\documentclass[12pt, twoside]{elsarticle}
\usepackage[english]{babel}
\usepackage{amsmath}
\usepackage{graphicx, color}
\usepackage{amssymb}

\newcommand{\so}[1]{\mbox{{\bf D}$^{2}\left\{#1\right \}$}}

\begin{document}

\begin{frontmatter}

\title{Theory of fission detector signals in reactor measurements -\\ detailed calculations}

\author{L. P\'al$^{1} $  and  I. P\'azsit$^{2}$}
\address{$^{1}$Centre for Energy Research, Hungarian Academy of Sciences, \\
H-1525 Budapest 114, POB 49, Hungary}
\address{$^{2}$Chalmers University of Technology,
Department of Applied Physics, \\
Division of Nuclear Engineering, \\SE-412 96 G\"oteborg, Sweden}

\begin{abstract}
The Campbell theorem, relating the variance of the current of a
fission chamber (a ``filtered Poisson process'') to the intensity of
the detection events and to the detector pulse shape, becomes
invalid when the neutrons generating the fission chamber current are
not independent. Recently a formalism was developed \cite{palpaz15}
by which the variance of the detector current could be calculated
for detecting neutrons in a subcritical multiplying system, where
the detection events are obviously not independent. In the present
paper, the previous formalism, which only accounted for prompt
neutrons, is generalised to account also for delayed neutrons. A
rigorous probabilistic analysis of the detector current was
performed by using the same simple, but realistic detector model as
in the previous work. The results of the present analysis made it
possible to determine the bias of the traditional Campbelling
techniques both qualitatively and quantitatively. The results show
that the variance still remains proportional to the detection
intensity, and is thus suitable for the monitoring of the mean flux,
but the calibration factor between the variance and the detection
intensity is an involved function of the detector pulse shape and
the subcritical reactivity of the system, which diverges for
critical systems.
\end{abstract}

\end{frontmatter}

\section{Introduction}

The main application area of fission chambers is the measurement of
the neutron flux in operating (critical) reactors \citep{filANE10}.
Fission chambers offer several advantages: they are robust; can be
operated in both pulse and current mode and they endure high
temperatures.

One special advantage of fission chambers is their capability of
suppressing unwanted minority components in the detector current,
such as gamma events, with a proper signal processing technique.
This is based on the so-called Campbell theorem, which establishes
relationships between the various order moments of the signal
\cite{papoulis65,pecseli00} , hence make it possible to determine
the mean detection rate from the second moment of the detector
current \cite{ezs15}.

However, using fission chambers in Campbelling mode in measurements
in a reactor has been controversial right from the beginning.
Namely, the Campbell theorems are only valid in the case when the
detection events are independent and the detection intervals obey an
exponential distribution. In the case of a multiplying system,
either critical or subcritical, the detections will not be
independent, due to the branching character of the neutron
multiplication process. In fact, it is the deviation of the
detection times from the exponential distribution, or the deviation
of the number of detections from a Poisson distribution, which is
utilised in the reactivity measurement techniques based on pulse
counting, i.e. the Feynman- and Rossi-alpha techniques
\cite{pazpal08}.

Even if it is surmised that application of the Campbell theorem
might be allowable even for detections of time-correlated neutrons,
by correcting for the (presumably small) quantitative error by a
calibration procedure, it is of fundamental importance to understand
the qualitative and quantitative effect of the existence of the
correlations between the detection events on the variance of the
signal. This would quantify the bias of the application of the
traditional Campbell relationships to extract the mean detection
intensity from the variance of the detector currents, and eventually
even make it possible to extract information about the system, such
as the subcritical reactivity, the same way as it is done by the
pulse counting techniques.

A first step to achieve these goals was made recently by the present
authors by setting up a formalism which unites the stochastic
description of the branching process with that of the statistical
theory of the detector signal \cite{palpaz15}. In that work, delayed
neutrons were not accounted for. In the present work we extend the
formalism to include also delayed neutrons. Naturally, the formalism
becomes more involved, and both the derivations and the results
become less transparent. Hence, out of the two goals of the previous
work (calculating the bias of the traditional Campbell theorem, and
calculating the auto-covariance of the detector signal in order to
determine the subcritical reactivity similarly as in a Rossi-alpha
measurement), only the first will be aimed at in this paper; the
second would lead to prohibitively complicated expressions.

One benefit of the more complicated calculations is that
quantitative results can be presented about the bias of the
traditional Campelling method in terms of the subcritical reactivity
of the system, with direct relevance to operating reactors. The
results show that the bias of the traditional Campbelling technique
is a monotonic function of the subcritical reactivity. It increases
when approaching criticality and diverges in a critical reactor. On
the other hand, in deep subcritical systems, the bias of the
traditional Campbelling method vanishes, and the traditional formula
becomes exact in the limit of a purely absorbing system where no
branching takes place and hence the individual detection events are
independent.

\section{Basic considerations}

The formalism used in this work was elaborated in two previous
publications. The statistical theory of the fission detector signal
due to independent detection events, based on the backward master
equation approach, was introduced in \cite{PPE14}. Then the theory
was extended to the case of detecting neutrons in a subcritical
medium driven with an extraneous neutron source with Poisson
statistics \cite{palpaz15}. All fission neutrons were considered as
prompt in this latter work. Although we stall try to make this paper
self-contained as much as possible, reference will be made to the
above publications for the details whenever it is practical.

As in the previous works, a basic quantity used will be the
probability
\begin{equation} \label{1}
{\mathcal P}\{y \leq \boldsymbol{\eta_{s}}(t) \leq y+dy \, \vert
\boldsymbol{\eta_{s}}(0)=0\} = h(y, t)\;dy + o(dy)
\end{equation}
that at a time instant $t \geq 0$ the value of a single detector
pulse $\boldsymbol{\eta_{s}}(t)$, initiated by the detection of a
neutron at time $t = 0$, is in the interval $[y, \, y + dy)$. We
assume that the current pulse generated by a neutron arriving to the
detector can be considered as a response function of the detector.
In many cases, this response function cannot be given by a
deterministic function $f(t)$; rather, it should be described by a
function $\varphi(\boldsymbol{\xi}, t)$ which depends on a properly
selected random variable $\boldsymbol{\xi}$. Hence the function
$\varphi(\boldsymbol{\xi}, t)$ represents the current signal which
exists in the detector at time $t \geq 0$ after the arrival of one
neutron at $t = 0$. We assume that this signal depends also on a
random variable $\boldsymbol{\xi}$, which is defined by its
distribution function $w(x)$
\begin{equation} \label{2}
{\mathcal P}\left\{\boldsymbol{\xi} \leq x \right\} =
\int_{-\infty}^{x} w(x')\;dx',
\end{equation}
thus one can write that
\begin{equation} \label{3}
h(y, t) = \int_{-\infty}^{+\infty}\,\delta\left[y - \varphi(x,
t)\right]\,w(x)\;dx, \qquad \qquad y \geq 0.
\end{equation}
The continuously arriving neutrons generate the detector current as
the aggregate of such current signals, each related to different
realizations of $\boldsymbol{\xi}$.

In further calculations the characteristic function
\begin{equation} \label{4}
\widetilde{h}(\omega, t) = \int_{-\infty}^{+\infty} e^{\imath \omega
y}\,h(y, t)\;dy
\end{equation}
will be often used, which is obtained from the above as
\begin{equation} \label{5}
\widetilde{h}(\omega, t) = \int_{-\infty}^{+\infty} e^{\imath \omega
y}\,\left(\int_{-\infty}^{+\infty}\,\delta\left[y - \varphi(x,
t)\right]\,w(x)\;dx \right)\;dy =
\int_{-\infty}^{+\infty}\,e^{\imath \omega \varphi(x, t)}\,w(x)\;dx.
\end{equation}
The moments of $\boldsymbol{\eta_{s}}(t)$ are given by the formulae:
\begin{equation} \label{5a}
\Phi_{k}(t) = \left(\frac{\displaystyle 1}{\displaystyle
\imath}\right)^{k}\, \left[\frac{\displaystyle
\partial^{k}\widetilde{h}(\omega, t)}{\displaystyle \partial
\omega^{k}}\right]_{\omega=0} = \left(\frac{\displaystyle
1}{\displaystyle \imath}\right)^{k} \, \int_{-\infty}^{+\infty}\,
\left[\varphi(x, t)\right]^{k}\,w(x)\;dx,
\end{equation}
\[ k = 1, 2, \ldots \;.\]
For any given signal shape $\varphi(x, t)$ and amplitude
distribution $w(x)$, these moments can be calculated.

The main objective of the present work is to determine the
stochastic properties of the detector signal for detections in a
multiplying system driven by a Poisson-like neutron source with
constant intensity, where the individual detection events are not
independent. In the previous work \cite{palpaz15} it was already
shown that the Campbell theorem becomes invalid when the detected
neutrons lose their independence. In the present work, the theory
will be extended for the case when the effect of the delayed
neutrons in the neutron multiplication process is taken into
account. For the sake of simplicity, only one type of precursors
will be considered in the calculations.

Denote by
\begin{equation} \label{6}
{\mathcal P}\left\{y \leq \boldsymbol{\eta}(t) \leq y+dy, t\, \vert 0\right\} =
P\left(y, t \, \vert 0\right)\;dy
\end{equation}
the probability that in a subcritical system which is driven by a
Poisson-like neutron source with constant intensity $s_{0}$, at the
time moment $t \geq 0$, the detector current $\boldsymbol{\eta}(t)$
is found in the interval $(y, \, y+dy]$, provided that at the time
instant $t=0$ the detector current and the numbers of neutrons and
precursors were zero. For further calculations one needs the
characteristic function
\begin{equation} \label{7}
G\left(\omega, t \, \vert 0\right) =
\int_{-\infty}^{+\infty} e^{\imath \omega y}\,P\left(y, t
\, \vert 0\right)\;dy,
\end{equation}
which will play the role of the generating function.

Adding up for the mutually exclusive events that there will be or
will not be a first collision between $[0,t]$ and applying the
convolution theorem for the latter case, one obtains the following
backward equation for $P\left(y, t \, \vert 0\right)$:
\[
P\left(y, t \, \vert 0\right) = e^{-s_{0}t}\,\delta(y) +
\]
\begin{equation} \label{9}
s_{0} \int_{0}^{t} e^{-s_{0}
(t-t')}\,\iint\limits_{y_{1}+y_{2}=y} p(y_{1}, t' \, \vert
\boldsymbol{n}(0) = 1)\,P\left(y_{2}, t' \, \vert 0\right)\;dy_{1} \,dy_{2},
\end{equation}
where
\[
p(y, t \, \vert \boldsymbol{n}(0) = 1)\;dy
\]
is the probability that in a subcritical system without a source, at
the time instant $t \geq 0$, the detector current
$\boldsymbol{\eta}(t)$ lies within the interval $(y, \, y+dy]$,
provided that at time $t=0$ the detector current and the number of
precursors were zero, while the number of the neutrons was equal to
$\textbf{1}$. One can call $p(y, t \, \vert \boldsymbol{n}(0) = 1)$
the single-particle induced distribution, whereas $P\left(y, t \,
\vert 0\right)$ is the source-induced distribution.

Introducing the characteristic function
\begin{equation} \label{11}
{g}(\omega,t  \, \vert \boldsymbol{n}(0) = 1) =
\int_{-\infty}^{+\infty} e^{\imath \omega y}\, {p}\left(y, t \,
\vert \boldsymbol{n}(0) = 1\right),
\end{equation}
and the convolution theorem, from equation (\ref{9}) one obtains the
following equation:
\begin{equation} \label{10}
{G}(\omega, t) = e^{-s_{0}t} + s_{0}
\int_{0}^{t} e^{-s_{0}(t-t')}\,{g}(\omega, t' \, \vert \boldsymbol{n}(0) = 1) \,
G(\omega, t')\;dt',
\end{equation}

It is easy to show that the solution of the integral equation
(\ref{10}) is given by
\begin{equation}  \label{12}
{G}(\omega, t) = \exp\left\{
s_{0}\left[\int_{0}^{t}{g}(\omega, t'  \, \vert \boldsymbol{n}(0) = 1) - 1 \right]\;dt'\right\}.
\end{equation}
It can be seen that the density function $p(y, t \, \vert
\boldsymbol{n}(0) = 1)$, or its characteristic function ${g}(\omega,
t' \, \vert \boldsymbol{n}(0) = 1)$, plays a fundamental role in the
description of random behaviour of the detector current.

It remains to derive an equation for $p(y, t \,\vert
\boldsymbol{n}(0) = 1)$, and it is at this point that our
description will deviate from the previous work. First of all, since
both the detector signal, as well as the number distribution of the
neutrons, will be the result of a branching process, one can only
derive a backward master equation if one also keeps track of the
time evolution of the number of neutrons. In addition, we need to
take into account that the branching will generate not only neutrons
but also delayed neutron precursors, as well as that a singe delayed
neutron precursor too, can initiate a branching process and a
corresponding detector signal evolution.

Hence, in deriving the corresponding master equations, we need to
consider the extended densities
\begin{equation}
p_{x}(y,n,c,t\, \vert \boldsymbol{n}(0) = 1)
\end{equation}
and
\begin{equation}
p_{x}(y,n,c,t\, \vert \boldsymbol{c}(0) = 1).
\end{equation}
Here, $p_{x}(y, n, c, t \, \vert \boldsymbol{n}(0) = 1)\;dy$ is the
probability that in a (source-free) subcritical system at the time
instant $t \geq 0$, the detector current $\boldsymbol{\eta}(t)$ will
lie within the interval $(y, \, y+dy]$, while the number of neutrons
$\boldsymbol{n}(t)$ and that of precursors $\boldsymbol{c}(t)$ are
equal to $n$ and $c$, respectively, provided that at the time
instant $t=0$ the detector current and the number of precursors was
zero, and the number of the neutrons was equal to $\textbf{1}$.
Likewise, $p_{x}(y, n, c, t \, \vert \boldsymbol{c}(0) = 1)\;dy$ is
the same probability, except that at time $t = 0$ the number of
neutrons was equal to zero, and the number of precursors was
$\textbf{1}$. Once these quantities are determined, the density
${p}\left(y, t \, \vert \boldsymbol{n}(0) = 1\right)$ appearing in
(\ref{9}) is obtained as
\begin{equation}
{p}\left(y, t \, \vert \boldsymbol{n}(0) = 1\right) =
\sum_{n=0}^{\infty}\,\sum_{c=0}^{\infty}\,p_{x}(y,n,c,t\, \vert
\boldsymbol{n}(0) = 1)
\end{equation}
and similarly for ${p}\left(y, t \, \vert \boldsymbol{c}(0) = 1\right)$.

The derivation of the backward equation for $p_{x}(y,n,c,t\, \vert
\boldsymbol{n}(0) = 1)$ goes as follows. If at $t=0$ one single
neutron exists in the system, then in the time interval $(0,\, t]$
four mutually exclusive events can take place:
\begin{itemize}
\item the neutron will not have any reaction;
\item the neutron gets detected in the detector
with an intensity $\lambda_{d}$ and creates a current pulse;
\item the neutron is captured in the subcritical medium with
intensity $\lambda_{c}$,
\item the neutron creates a fission in the subcritical medium
with intensity $\lambda_{f}$.
\end{itemize}

\noindent It is clear that the total intensity of a reaction in the
system is $\lambda_{r} = \lambda_{d} + \lambda_{c} + \lambda_{f}$.
Similarly to previous work, the fact that the detection itself is a
fission event, which also will produce further neutrons, will be
neglected.

\vspace{0.2cm}

In a fission reaction, $k \geq 0$ neutrons and $\ell \geq 0$
precursors of the same type are produced with probability $f(k,
\ell)$. It is assumed that the number of neutrons and that of
precursors are independent, i.e.
\begin{equation}  \label{13}
f(k, \ell) = f_{k}^{(p)}\, f_{\ell}^{(d)}.
\end{equation}
For later use, introduce the generating functions
\begin{equation} \label{14}
q^{(p)}(z) = \sum_{k=0}^{\infty} f_{k}^{(p)} \, z^{k} \qquad
\text{and} \qquad q^{(d)}(z) = \sum_{\ell=0}^{\infty} f_{\ell}^{(d)}
\, z^{\ell}.
\end{equation}
By applying the backward approach, one can write
\[p_{x}(y, n, c, t \, \vert \boldsymbol{n}(0) = 1) = e^{-\lambda_{r}
t}\,\delta(y)\,\delta_{n,1}\,\delta_{c,0} +
\lambda_{d}\,\delta_{n,0}\,\delta_{c,0}\,\int_{0}^{t} e^{-\lambda_{r}
(t-t')}\,h(y, t')\;dt' + \]
\[\lambda_{c}\,\delta_{n,0}\,\delta_{c,0}\,\int_{0}^{t} e^{-\lambda_{r}
(t-t')}\;dt' + \]
\[\lambda_{f}\,\int_{0}^{t} e^{-\lambda_{r}
(t-t')}\,\sum_{k}\,\sum_{\ell}\,f^{(p)}(k)\,f^{(d)}(\ell) \,
\iint\limits_{y_{1}+y_{2}=y} \, \sum\limits_{n_{1}+n_{2}=n} \,
\sum\limits_{c_{1}+c_{2}=c} \, \times
\]
\begin{equation} \label{15}
U_{k}(y_{1}, n_{1}, c_{1}, t' \, \vert \boldsymbol{n}(0) = 1)\,
V_{\ell}(y_{2}, n_{2}, c_{2}, t' \, \vert \boldsymbol{c}(0) = 1)
dy_{1}\,dy_{2}\,dt',
\end{equation}
where
\[ U_{k}(y_{1}, n_{1}, c_{1}, t' \, \vert \boldsymbol{n}(0) = 1) =
\left[1-\Delta(k)\right] \, \delta(y_{1})
\,\delta_{n_{1},0}\,\delta_{c_{1},0} +
\]
\begin{equation} \label{16}
\Delta(k) \idotsint\limits_{y_{11}+\cdots + y_{1k}=y_{1}} \,
\sum\limits_{n_{11}+\cdots n_{1k}=n_{1}} \,
\sum\limits_{c_{11}+\cdots c_{1k}=c_{1}}\,
\prod_{j=1}^{k}\,p_{x}(y_{1j}, n_{1j}, c_{1j}, t' \, \vert
\boldsymbol{n}(0) = 1)\;dy_{1j}
\end{equation}
and
\[ V_{\ell}(y_{2}, n_{2}, c_{2}, t' \, \vert \boldsymbol{c}(0) = 1) =
\left[1-\Delta(\ell)\right] \, \delta(y_{2})
\,\delta_{n_{2},0}\,\delta_{c_{2},0} +
\]
\begin{equation} \label{17}
\Delta(\ell) \idotsint\limits_{y_{21}+\cdots + y_{2\ell}=y_{2}} \,
\sum\limits_{n_{21}+\cdots n_{2\ell}=n_{2}} \,
\sum\limits_{c_{21}+\cdots c_{2\ell}=c_{2}}\,
\prod_{j=1}^{\ell}\,p_{x}(y_{2j}, n_{2j}, c_{2j}, t' \, \vert
\boldsymbol{c}(0) = 1)\;dy_{2j}.
\end{equation}
In a similar manner, taking into account the two mutually exclusive events that the delayed neutron precursor will not decay or will decay with intensity $\lambda$, the following equation can be derived for the case when the branching process is started by one precursor:
\begin{equation} \label{18}
p_{x}(y, n, c, t \, \vert \boldsymbol{c}(0) = 1) = e^{-\lambda
t}\,\delta(y)\,\delta_{n,0}\,\delta_{c,1} + \lambda\,\int_{0}^{t}
e^{-\lambda (t-t')} \, p_{x}(y, n, c, t' \, \vert \boldsymbol{n}(0)
= 1) \;dt',
\end{equation}
which connects the density function $p_{x}(y, n, c, t \, \vert
\boldsymbol{c}(0) = 1)$ with $p_{x}(y, n, c, t \, \vert
\boldsymbol{n}(0) = 1)$.

Defining the generating functions
\begin{equation}
g_{x}(\omega, z_{1}, z_{2}, t \, \vert \boldsymbol{n}(0) = 1) =
\int_{-\infty}^{+\infty} e^{\imath \omega y}
\sum_{n=0}^{\infty}\,\sum_{c=0}^{\infty}\,p_{x}(y, n, c, t \, \vert
\boldsymbol{n}(0) = 1)\,z_{1}^{n}\,z_{2}^{c}\; dy
\end{equation}
and
\begin{equation}
g_{x}(\omega, z_{1}, z_{2}, t \, \vert \boldsymbol{c}(0) = 1) =
\int_{-\infty}^{+\infty} e^{\imath \omega y}
\sum_{n=0}^{\infty}\,\sum_{c=0}^{\infty}\,p_{x}(y, n, c, t \, \vert
\boldsymbol{c}(0) = 1)\,z_{1}^{n}\,z_{2}^{c}\; dy,
\end{equation}
from (\ref{15}) one obtains the equations for the generating
functions in the following form:
\[
g_{x}(\omega, z_{1}, z_{2}, t \, \vert
\boldsymbol{n}(0) = 1) = e^{- \lambda_{r} t}\, z_{1} +
\lambda_{d}\,\int_{0}^{t} e^{-\lambda_{r} (t-t')}\,
\widetilde{h}(\omega, t')\;dt' + \lambda_{c}\,\int_{0}^{t}
e^{-\lambda_{r} (t-t')} \;dt' +
\]
\begin{equation} \label{19}
\lambda_{f}\,\int_{0}^{t} e^{-\lambda_{r} (t-t')}
q^{(p)}\left[g_{x}(\omega, z_{1}, z_{2}, t' \, \vert
\boldsymbol{n}(0) = 1)\right] \,
q^{(d)}\left[g_{x}(\omega, z_{1}, z_{2}, t' \, \vert
\boldsymbol{c}(0) = 1)\right],
\end{equation}
while the generating function of  equation (\ref{18}) is given by
\begin{equation} \label{20}
g_{x}(\omega, z_{1}, z_{2}, t  \, \vert
\boldsymbol{c}(0) = 1) = e^{ - \lambda t} \, z_{2} +
\lambda\,\int_{0}^{t} e^{-\lambda (t-t')} \,
g_{x}(\omega, z_{1}, z_{2}, t' \, \vert
\boldsymbol{n}(0) = 1) \;dt'.
\end{equation}
At this point it is possible to revert to the case when only the
stochastic behaviour of the detector current is of interest only,
irrespective of the number of the neutrons or precursors in the
system. The dynamics of the branching is compressed into the
non-linear functions $q^{(p)}\left[\dots \right]$ and
$q^{(d)}\left[\dots \right]$. Hence in the continuation we will
simplify Eqs (\ref{19}) and (\ref{20}) by substituting $z_{1} =
z_{2} =1$ and by turning to the quantities
\begin{equation}\label{21a}
g(\omega, t \, \vert \boldsymbol{n}(0) = 1) = g_{x}(\omega, 1, 1, t
\, \vert \boldsymbol{n}(0) = 1)
\end{equation}
and
\begin{equation} \label{21}
g(\omega, t \, \vert \boldsymbol{c}(0) = 1) = g_{x}(\omega, 1, 1, t
\, \vert \boldsymbol{c}(0) = 1)
\end{equation}
As is seen from (\ref{12}) it is the $g(\omega, t \, \vert
\boldsymbol{n}(0) = 1)$ of (\ref{21a}) which is needed for the
calculation of ${G}(\omega, t)$, from which the moments of the
stationary detector current in a subcritical system driven by an
external neutron noise can be determined.

From equations (\ref{19}) and (\ref{20}), applying the notations
defined in (\ref{21a}) and (\ref{21}), one obtains
\[
g(\omega, t \, \vert \boldsymbol{n}(0) = 1) =
e^{- \lambda_{r} t} + \lambda_{d}\,\int_{0}^{t} e^{-\lambda_{r}
(t-t')} \, \widetilde{h}(\omega, t')\;dt' +
\lambda_{c}\,\int_{0}^{t} e^{-\lambda_{r} (t-t')} \;dt' + \]
\begin{equation} \label{23}
\lambda_{f}\,\int_{0}^{t} e^{-\lambda_{r} (t-t')}
q^{(p)}\left[g(\omega, t' \, \vert \boldsymbol{n}(0) =
1)\right] \, q^{(d)}\left[g(\omega, t' \, \vert
\boldsymbol{c}(0) = 1)\right],
\end{equation}
and
\begin{equation} \label{24}
g(\omega, t \, \vert \boldsymbol{c}(0) = 1) = e^{ - \lambda t}
+ \lambda\,\int_{0}^{t} e^{-\lambda (t-t')} \, g(\omega, t' \,
\vert \boldsymbol{n}(0) = 1) \;dt'.
\end{equation}

For the determination of the cumulants of the detector current we will
use the well-known relation
\begin{equation} \label{25}
\kappa_{n}(t) = \left(\frac{\displaystyle 1}{\displaystyle
\imath}\right)^{n}\,\left[\frac{\displaystyle
\partial^{n}\,\mathbb{K}(\omega, t)}{\displaystyle
\partial \omega^{n}}\right]_{\omega=0},
\end{equation}
where
\begin{equation} \label{26}
\mathbb{K}(\omega, t) = \ln G(\omega, t) =
s_{0}\left[\int_{0}^{t}g(\omega, t \, \vert \boldsymbol{n}(0)
= 1) - 1\right]\;dt'.
\end{equation}

\section{Expectation of the detector current}

By using expressions (\ref{25}) and (\ref{26}), one can write
\begin{equation} \label{27}
\left<\boldsymbol{\eta}(t)\right> = \kappa_{1}(t) \equiv I_{1}(t) =
s_{0} \, \frac{\displaystyle 1}{\displaystyle
\imath}\,\int_{0}^{t}\left[\frac{\displaystyle
\partial g(\omega, t' \, \vert \boldsymbol{n}(0) = 1)}
{\displaystyle \partial \omega}\right]_{\omega=0} \;dt',
\end{equation}
where
\begin{equation} \label{28}
\frac{\displaystyle 1}{\displaystyle
\imath}\,\left[\frac{\displaystyle
\partial g(\omega, t \, \vert \boldsymbol{n}(0) = 1)}
{\displaystyle \partial \omega}\right]_{\omega=0} = i_{1}(t \,\vert
\boldsymbol{n}(0) = 1)
\end{equation}
is the expectation of the detector current generated by a single
starting neutron. From (\ref{23}) one can derive the equation
\[i_{1}(t \,\vert \boldsymbol{n}(0) = 1) = \lambda_{d}\,\int_{0}^{t}
e^{-\lambda_{r} (t-t')} \, \Phi_{1}(t') \;dt' + \]
\begin{equation} \label{29}
\lambda_{f}\,\int_{0}^{t} e^{-\lambda_{r} (t-t')}
\left[q_{1}^{(p)}\, i_{1}(t' \, \vert \boldsymbol{n}(0) = 1) +
q_{1}^{(d)}\, i_{1}(t' \, \vert \boldsymbol{c}(0) = 1) \right]
\;dt',
\end{equation}
where
\begin{equation} \label{30}
\Phi_{1}(t') = \frac{\displaystyle 1}{\displaystyle \imath}
\,\left[\frac{\displaystyle \partial \widetilde{h}(\omega,
t')}{\displaystyle \partial \omega}\right]_{\omega=0},
\end{equation}
and
\[q_{1}^{(p)} = \left[\frac{\displaystyle dq^{(p)}(z)}{\displaystyle
dz}\right]_{z=1} = \nu_{p}, \qquad \text{while} \qquad q_{1}^{(d)}
= \left[\frac{\displaystyle dq^{(d)}(z)}{\displaystyle
dz}\right]_{z=1} = \nu_{d}.\]

In order to obtain the solution of (\ref{29}), one has to take into
account the relation
\begin{equation} \label{31}
i_{1}(t \, \vert \boldsymbol{c}(0) = 1) = \lambda\,\int_{0}^{t}
e^{-\lambda (t-t')} \, i_{1}(t' \, \vert \boldsymbol{n}(0) = 1)
\;dt'
\end{equation}
which follows from (\ref{24}), and apply the Laplace transforms
\begin{equation} \label{32}
\widetilde{i}_{1}(s \, \vert \boldsymbol{n}(0) = 1) = \int_{0}^{t}
e^{-s t} \, i_{1}(t \, \vert \boldsymbol{n}(0) = 1)\; dt
\end{equation}
and
\begin{equation} \label{33}
\widetilde{i}_{1}(s \, \vert \boldsymbol{c}(0) = 1) = \int_{0}^{t}
e^{-s t} \, i_{1}(t \, \vert \boldsymbol{c}(0) = 1) \; dt,
\end{equation}
as well as
\begin{equation} \label{33a}
\widetilde{\Phi}_{1}(s) = \int_{0}^{t} e^{-s t} \, \Phi_{1}(t)\;dt.
\end{equation}
It is seen that the Laplace transform of (\ref{29})
satisfies the equation:
\[ \widetilde{i}_{1}(s \,\vert \boldsymbol{n}(0) = 1) = \]
\begin{equation} \label{34}
\frac{\displaystyle \lambda_{d}}{\displaystyle s + \lambda_{d}
+\lambda_{c} + \lambda_{f}} \,\widetilde{\Phi}_{1}(s) +
\frac{\displaystyle \lambda_{f}}{\displaystyle s + \lambda_{d} +
\lambda_{c} + \lambda_{f}}\, \left[\nu_{0}\,
\widetilde{i}_{1}(s \, \vert \boldsymbol{n}(0) = 1) +
\nu_{d} \, \widetilde{i}_{1}(s \, \vert \boldsymbol{c}(0) = 1)
\right],
\end{equation}
while the Laplace transform of (\ref{31}) obeys the equation
\begin{equation} \label{35}
\widetilde{i}_{1}(s \, \vert \boldsymbol{c}(0) = 1) =
\frac{\displaystyle \lambda}{\displaystyle s+\lambda}
\,\widetilde{i}_{1}(s \, \vert \boldsymbol{n}(0) = 1).
\end{equation}
After elementary algebra, from Eqs (\ref{34}) and (\ref{35}), one
obtains the Laplace transform of the expectation of the detector
current generated by a single starting neutron in the following
form:
\begin{equation}  \label{36}
\widetilde{i}_{1}(s \,\vert \boldsymbol{n}(0) = 1) =
\frac{\displaystyle \lambda_{d}\,(s + \lambda) \,
\widetilde{\Phi}_{1}(s)}{\displaystyle (s + \lambda )\,\left[s
+\lambda_{d} + \lambda_{c} + \lambda_{f}(1 - \nu_{0})\right] -
\lambda \, \lambda_{f}\, \nu_{d}}.
\end{equation}
By using conventional notations, one has
\begin{equation}  \label{37}
\widetilde{i}_{1}(s \,\vert \boldsymbol{n}(0) = 1) =
\frac{\displaystyle \lambda_{d}\,(s + \lambda) \,
\widetilde{\Phi}_{1}(s)}{\displaystyle (s + \lambda)\,\left(s +
\frac{\displaystyle \beta - \rho}{\displaystyle \Lambda}\right) -
\lambda \,\frac{\displaystyle \beta }{\displaystyle \Lambda}},
\end{equation}
where
\begin{equation} \label{38}
\rho = \frac{\displaystyle  \nu \,\lambda_{f} - \left(\lambda_{d} +
\lambda_{c} + \lambda_{f}\right)}{\displaystyle \nu \,\lambda_{f}}
\end{equation}
is the reactivity, while $\nu = \nu_{p} + \nu_{d}$ and $ \beta =
\frac{\displaystyle \nu_{d}}{\displaystyle \nu}$. Further, the
following notations will also be used:
\[\Lambda = \frac{\displaystyle 1}{\displaystyle \nu \,\lambda_{f}}, \]
which is the prompt neutron generation time and
\begin{equation}\label{alpha}
\alpha =
\frac{\displaystyle \beta - \rho}{\displaystyle \Lambda} > 0
\end{equation}
is the prompt neutron decay constant.

In terms of the negative roots $s_{1}$ and $s_{2}$ of the
characteristic equation
\[
s^{2} + s\,\left(\lambda+ \alpha\right) - \frac{\displaystyle
\lambda \rho}{\displaystyle \Lambda} = 0,
\]
expression (\ref{37}) can be rewritten in the following form:
\begin{equation} \label{39}
\widetilde{i}_{1}(s \,\vert \boldsymbol{n}(0) = 1) =
\frac{\displaystyle \lambda_{d}\,(s + \lambda) \,
\widetilde{\Phi}_{1}(s)}{\displaystyle (s + s_{1})\,(s + s_{2})},
\end{equation}
where
\begin{equation} \label{40}
s_{1} = \frac{\displaystyle 1}{\displaystyle 2}\,\left[\lambda +
\alpha +
\sqrt{\left(\lambda + \alpha \right)^{2} + 4\,\frac{\displaystyle
\lambda\,\rho}{\displaystyle \Lambda}}\right],
\end{equation}
and
\begin{equation} \label{41}
s_{2} = \frac{\displaystyle 1}{\displaystyle 2}\,\left[\lambda +
\alpha -
\sqrt{\left(\lambda + \alpha \right)^{2} + 4\,\frac{\displaystyle
\lambda\,\rho}{\displaystyle \Lambda}}\right].
\end{equation}

The expectation of the detector current, generated by a chain of
neutrons generated in a subcritical multiplying assembly, started by
one single source neutron, can be obtained by the inverse Laplace
transformation of (\ref{39}). It is easy to show that
\begin{equation} \label{42}
i_{1}(t \,\vert \boldsymbol{n}(0) = 1) = \lambda_{d}\, \int_{0}^{t}
\, \frac{\displaystyle (s_{1} - \lambda)\,e^{- s_{1} (t-u)} - (s_{2}
- \lambda)\,e^{- s_{2} (t-u)}}{\displaystyle  s_{1} -
s_{2}}\,\Phi_{1}(u)\;du.
\end{equation}
In order to obtain the expectation of the detector current in a
subcritical multiplying medium driven by a stationary Poisson source
of neutrons with intensity $s_{0}$ , one has to calculate the
integral
\begin{equation} \label{43}
I_{1}(t) = s_{0}\, \int_{0}^{t} i_{1}(t' \,\vert \boldsymbol{n}(0) =
1)\;dt',
\end{equation}
the Laplace transform of which is given by the formula
\begin{equation} \label{44}
\widetilde{I}_{1}(s) = s_{0}\, \frac{\displaystyle \lambda_{d}\,(s +
\lambda) \, \widetilde{\Phi}_{1}(s)}{\displaystyle  s\, (s +
s_{1})\,(s + s_{2})}.
\end{equation}
By using the Tauber theorem, one can determine the asymptotically
stationary expectation of the detector current. Since
\begin{equation} \label{45}
\lim_{t \rightarrow \infty}\, I_{1}(t) = I_{1}^{(st)} = \lim_{s
\rightarrow 0}\,s\,\widetilde{I}_{1}(s) = s_{0}\,
\frac{\displaystyle \lambda_{d} \, \lambda \,
\widetilde{\Phi}_{1}(0)}{\displaystyle s_{1} \, s_{2}} = s_{0}\,
\frac{\displaystyle \lambda_{d} \,
\widetilde{\Phi}_{1}(0)}{\displaystyle -\rho/\Lambda},
\end{equation}
it is obvious that a stationary expectation of the detector current
exists only when $\rho \leq 0$, i.e. when the multiplying assembly
is in a subcritical state. The quantity $\widetilde{\Phi}_{1}(0)$
characterizes the average value of the electrical charge produced in
the detector during the registration of one neutron.

\subsection{A concrete example for the expectation}

Eqs (\ref{44}) and (\ref{45}) show that the expression for the
expectation of the  detector current contains the average current
pulse $\Phi_{1}(t)$, generated by the detection of a single neutron
at time $t=0$. In general, the shape of $\Phi_{1}(t)$ depends on a
number of physical processes taking place in the detector during the
rather complicated processes of charge generation and transport.
However, in the present work, like in its predecessors, we will only
account for the fluctuations of the detector current due to the
randomness of the arrival times of the neutrons. Therefore,
similarly to Refs \cite{PPE14} and \cite{palpaz15}, we will choose a
constant value $\alpha_{e}$ instead of the random variable
$\boldsymbol{\xi}$, defined by (\ref{2}), i.e. we will use the
density function $w(x) = \delta(x - \alpha_{e})$ in the formula
(\ref{5a}). From this it follows that
\begin{equation} \label{46}
\Phi_{1}(t) = \varphi(\alpha_{e}, t),
\end{equation}
Based on the shapes of experimentally observed current pulses it
appears that the empirical expression
\begin{equation} \label{46a}
\Phi_{1}(t) = \varphi(\alpha_{e}, t) =
\alpha_{e}^{2}\,t\,e^{-\alpha_{e}\,t}\,Q
\end{equation}
is an acceptable approximation, hence it will be used for our
illustrative calculations. It is seen that $\alpha_{e}$ plays the
role of the decay constant of the detector pulse, and $Q$ is the
mean value of the charge collected in the case of the detection a
single neutron.
\begin{figure}[ht!]
\protect \centering{
\includegraphics[scale=0.7] {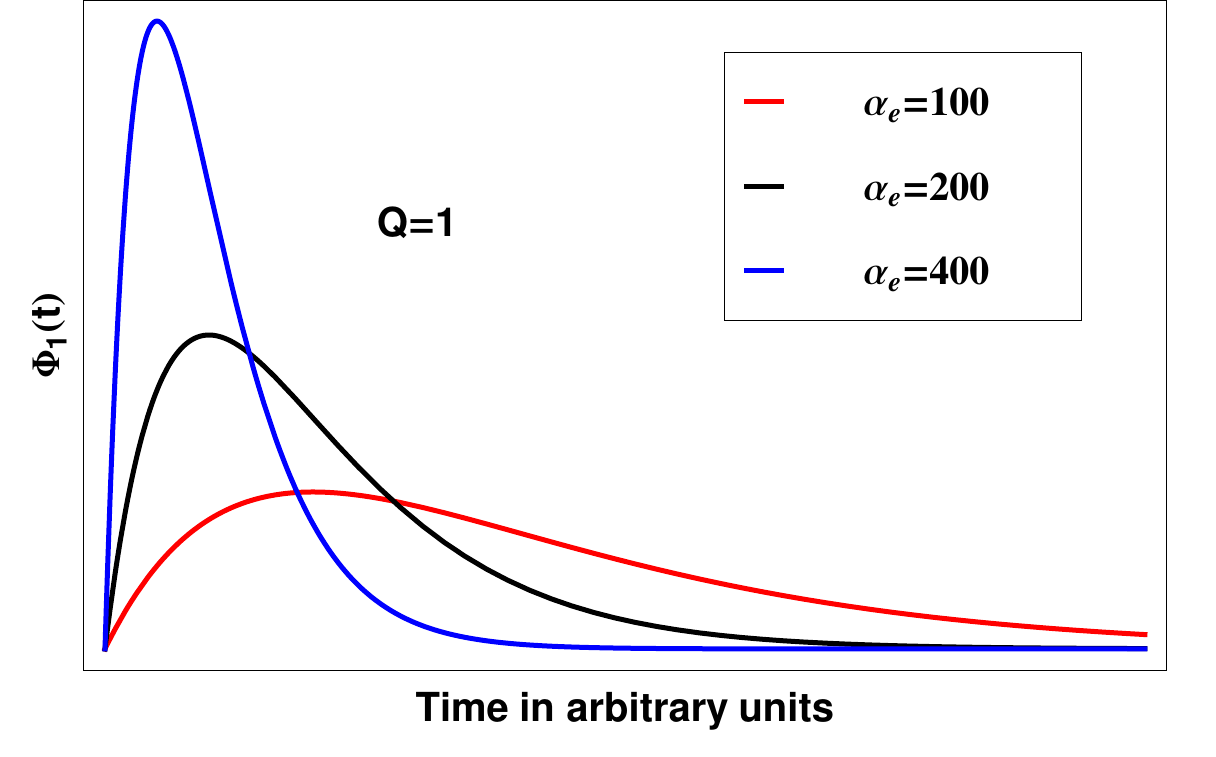}}\protect
\caption{\label{fig1}\footnotesize{Time dependence of a single
average current pulse }}
\end{figure}
\begin{figure}[ht!]
\protect \centering{
\includegraphics[scale=0.7] {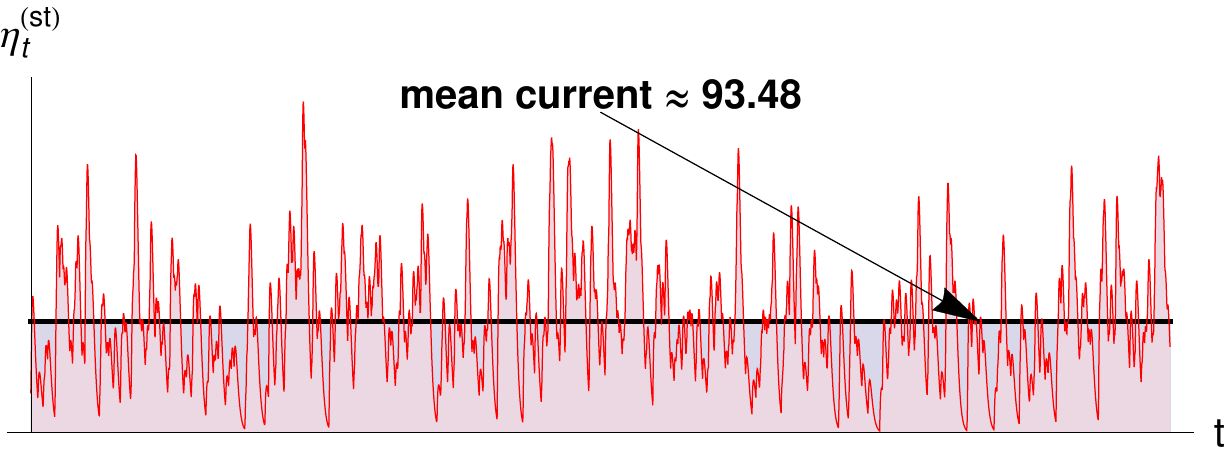}}\protect
\caption{\label{fig2}\footnotesize{Stationary detector current
$\boldsymbol{\eta}_{t}^{(st)}$ in a given time interval}}
\end{figure}

Fig. \ref{fig1} illustrates the shape of the time dependence of a
single average current pulse for various values of $\alpha_{e}$,
whereas Fig. \ref{fig2} shows the stationary detector current
$\boldsymbol{\eta}_{t}^{(st)}$ during a time interval for an
aggregate of several pulses. Obviously, the expectation and the
variance of the  $\boldsymbol{\eta}_{t}^{(st)}$ are constant.

In order to evaluate (\ref{39}) with the $\Phi(t)$ of (\ref{46}),
one needs the Laplace transform of $\Phi_{1}(t)$, which is obtained
as
\begin{equation} \label{47}
\widetilde{\Phi}_{1}(s) = \int_{0}^{\infty} e^{-s t} \,
\Phi_{1}(t)\;dt = \left(\frac{\displaystyle
\alpha_{e}}{\displaystyle s + \alpha_{e}}\right)^{2}\,Q.
\end{equation}
Hence one arrives at
\begin{equation}  \label{48}
\widetilde{i}_{1}(s \,\vert \boldsymbol{n}(0) = 1) =
\frac{\displaystyle \lambda_{d}\,(s + \lambda)}{\displaystyle (s +
s_{1})\,(s + s_{2})}\,\left(\frac{\displaystyle
\alpha_{e}}{\displaystyle s + \alpha_{e}}\right)^{2}\,Q,
\end{equation}
where $s_{1}$ and $s_{2}$ are defined by (\ref{40}) and (\ref{41}),
respectively. The inverse Laplace transform of (\ref{48}) is
obtained in the form
\[i_{1}(t \,\vert \boldsymbol{n}(0) = 1) =
\lambda_{d}\,\alpha_{e}^{2}\,Q\;\left[\frac{\displaystyle s_{1} -
\lambda}{\displaystyle \left(s_{1} - s_{2}\right)\,\left(s_{1} -
\alpha_{e}\right)^{2}}\, e^{ - s_{1} t} - \frac{\displaystyle s_{2}
- \lambda}{\displaystyle \left(s_{1} - s_{2}\right)\,\left(s_{2} -
\alpha_{e}\right)^{2}}\, e^{ - s_{2} t} \right] + \]
\[ \lambda_{d}\,\alpha_{e}^{2}\,Q\;\left\{\frac{\displaystyle  \left(2
- s_{2} t\right)\,\alpha_{e}\,\lambda - \alpha_{e}^{3}\,t -
s_{2}\,\lambda + \alpha_{e}^{2}\left(-1 + s_{2}\,t + \lambda \,
t\right)}{\left(s_{1} - \alpha_{e}\right)^{2}\,\left(s_{2} -
\alpha_{e}\right)^{2}} + \right.\]
\begin{equation} \label{49}
\left. \frac{\displaystyle
s_{1}\left\{\alpha_{e}\,t\,\left(\alpha_{e} - \lambda\right) -
\lambda + s_{2}\left[1 + t\,\left(-\alpha_{e} +
\lambda\right)\right]\right\}}{\left(s_{1} -
\alpha_{e}\right)^{2}\,\left(s_{2} -
\alpha_{e}\right)^{2}}\right\}\,e^{- \alpha_{e} t}.
\end{equation}

\begin{figure}[ht!]
\protect \centering{
\includegraphics[scale=0.75] {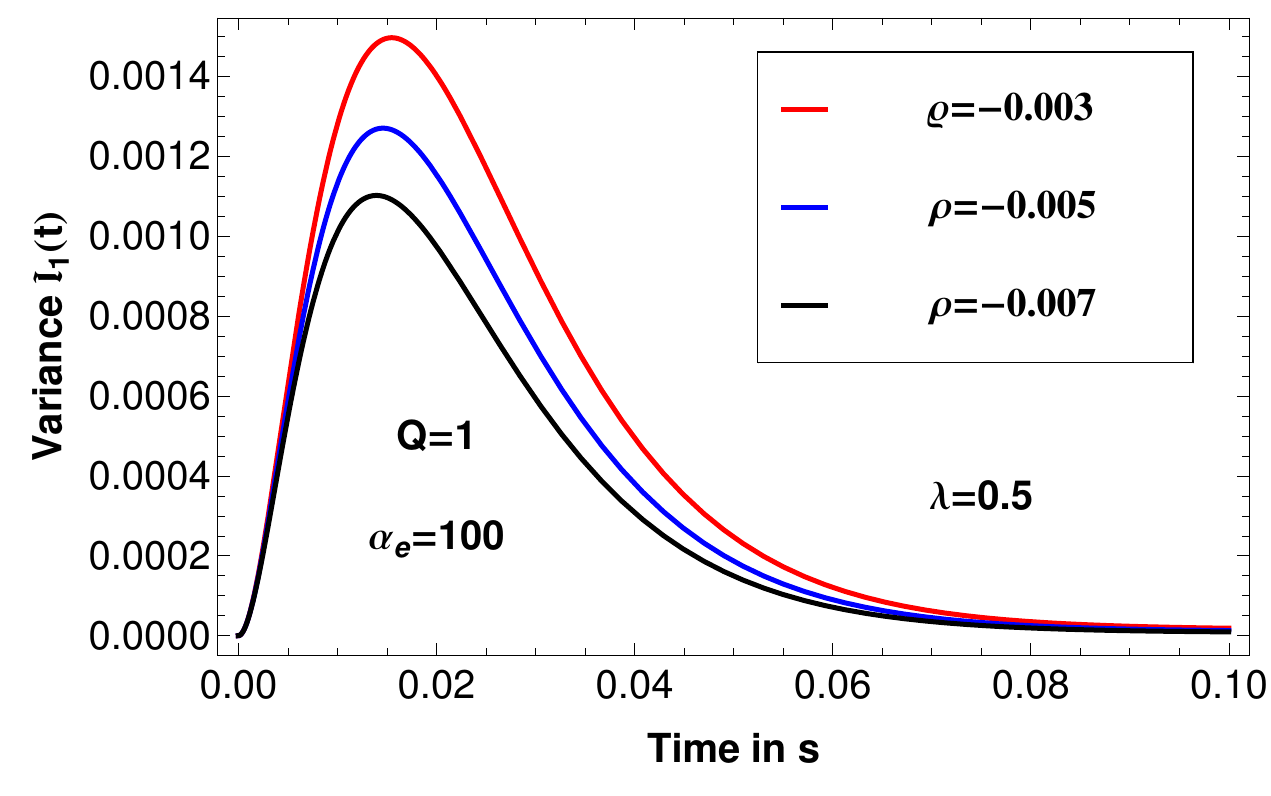}}\protect
\caption{\label{fig3}\footnotesize{Time dependence of the
expectation of the detector signal due to the reaction chain induced
in a subcritical assembly by one single starting neutron, for three
different reactivities }}
\end{figure}

\begin{figure}[ht!]
\protect \centering{
\includegraphics[scale=0.75] {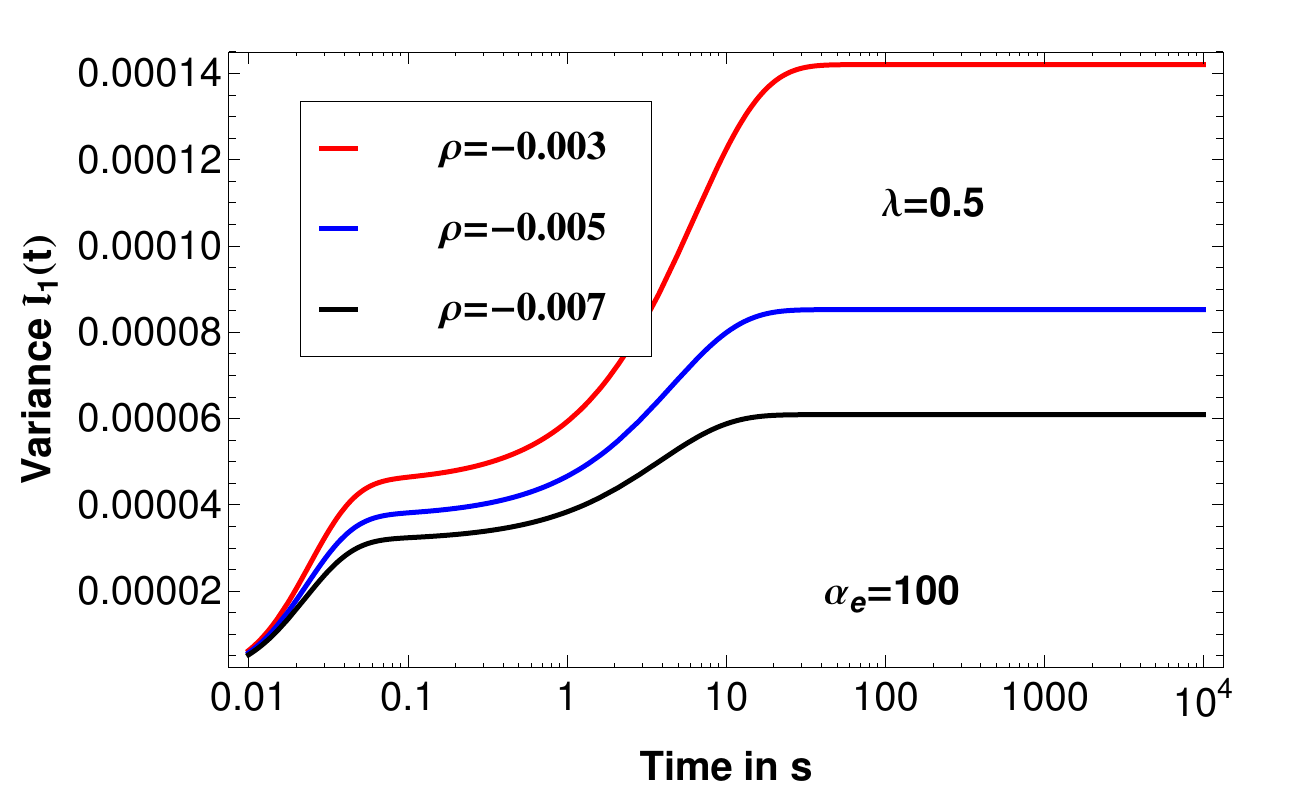}}\protect
\caption{\label{fig4}\footnotesize{Approach to the stationary value
of the expectation of the detector current in a subcritical assembly
with a source switched on at $t=0$, for three different
reactivities}}
\end{figure}
Fig. \ref{fig3} displays (\ref{49}), showing the time dependence of
the expectation of the detector signal due to the neutron chains
induced in a subcritical assembly by one single starting neutron,
for three different reactivities.

The time evolution of the detector current from time $t=0$, at which
time the external source was switched on in a system that previously
did not contain any neutrons, is given by
\begin{equation} \label{50}
I_{1}(t) = s_{0}\,\int_{0}^{t} i_{1}(t' \,\vert \boldsymbol{n}(0) =
1)\;dt'.
\end{equation}

Fig. \ref{fig4} shows the expected effect of the delayed neutrons,
forming an intermediate plateau-like part in the curves approaching
the stationary mean value of the detector current. In order to
calculate the the stationary expectation of the detector current,
i.e. the quantity
\begin{equation} \label{51}
I_{1}^{(st)} = \lim_{t \rightarrow \infty}\,I_{1}(t),
\end{equation}
one can use the Tauber theorem as in (\ref{45}). Taking into account that
$\widetilde{\Phi}_{1}(0 )=Q $, one has
\begin{equation} \label{52}
I_{1}^{(st)} = s_{0}\,\frac{\displaystyle \lambda_{d}}
{-\rho/\Lambda} \, Q, \qquad \text{where} \qquad \rho = \beta -
\alpha\,\Lambda < 0.
\end{equation}

\section{Variance of the detector current}

By using (\ref{25}) and (\ref{26}), one can write the variance of
the detector current for $t \geq 0$ in the form:
\begin{equation} \label{53}
\so{\boldsymbol{\eta}}  = \kappa_{2}(t) \equiv I_{2}(t) = s_{0} \,
\left(\frac{\displaystyle 1}{\displaystyle \imath}\right)^{2}
\,\int_{0}^{t}\left[\frac{\displaystyle
\partial^{2} g(\omega, t' \, \vert \boldsymbol{n}(0) = 1)}
{\displaystyle \partial \omega^{2}}\right]_{\omega=0} \;dt',
\end{equation}
where
\begin{equation} \label{54}
\left(\frac{\displaystyle 1}{\displaystyle \imath}\right)^{2}
\,\left[\frac{\displaystyle
\partial^{2} g(\omega, t' \, \vert \boldsymbol{n}(0) = 1)}
{\displaystyle \partial \omega^{2}}\right]_{\omega=0} = i_{2}(t'
\,\vert \boldsymbol{n}(0) = 1)
\end{equation}
is the variance of the detector current generated by a single
starting neutron. An equation for $i_{2}(t \,\vert \boldsymbol{n}(0)
= 1)$ can be derived from (\ref{23}). After a simple algebra one
obtains
\[
i_{2}(t \,\vert \boldsymbol{n}(0) = 1) =
\]
\begin{equation} \label{55}
\lambda_{d}\,\int_{0}^{t} e^{-\lambda_{r} (t-t')} \, \Phi_{2}(t')
\;dt' + \lambda_{f}\,\int_{0}^{t} e^{-\lambda_{r} (t-t')}
\left\{\left(\frac{\displaystyle 1}{\displaystyle
\imath}\right)^{2}\,\left[\frac{\displaystyle
\partial^{2} q(\omega, t')}
{\displaystyle \partial \omega^{2}}\right]_{\omega=0} \right\}\;dt',
\end{equation}
where
\begin{equation} \label{56}
\Phi_{2}(t') = \left(\frac{\displaystyle 1}{\displaystyle
\imath}\right)^{2} \,\left[\frac{\displaystyle \partial^{2}
\widetilde{h}(\omega, t')}{\displaystyle \partial
\omega^{2}}\right]_{\omega=0},
\end{equation}
and
\begin{equation} \label{57}
q(\omega, t') = q^{(p)}\left[g(\omega, t' \, \vert
\boldsymbol{n}(0) = 1)\right]\,q^{(d)}\left[g(\omega, t' \,
\vert \boldsymbol{c}(0) = 1)\right].
\end{equation}
In the first step one writes
\[\left(\frac{\displaystyle 1}{\displaystyle
\imath}\right)^{2}\,\left[\frac{\displaystyle
\partial^{2} q(\omega, t')}
{\displaystyle \partial \omega^{2}}\right]_{\omega=0} =
\left(\frac{\displaystyle 1}{\displaystyle
\imath}\right)^{2}\,\left\{\frac{\displaystyle \partial^{2}
q^{(p)}\left[g(\omega, t' \, \vert \boldsymbol{n}(0) =
1)\right]}{\displaystyle \partial \omega^{2}}\right\}_{\omega=0} +
\] \[2 \, q_{1}^{(p)}\, q_{1}^{(d)}\,i_{1}(t' \,\vert \boldsymbol{n}(0) =
1) \, i_{1}(t' \,\vert \boldsymbol{c}(0) = 1) +
\left(\frac{\displaystyle 1}{\displaystyle
\imath}\right)^{2}\,\left\{\frac{\displaystyle \partial^{2}
q^{(d)}\left[g(\omega, t' \, \vert \boldsymbol{c}(0) =
1)\right]}{\displaystyle \partial \omega^{2}}\right\}_{\omega=0}, \]
where
\[
\left(\frac{\displaystyle 1}{\displaystyle
\imath}\right)^{2}\,\left\{\frac{\displaystyle \partial^{2}
q^{(p)}\left[g(\omega, t' \, \vert \boldsymbol{n}(0) =
1)\right]}{\displaystyle \partial \omega^{2}}\right\}_{\omega=0} =
q_{2}^{(p)}\,\left[i_{1}(t' \,\vert \boldsymbol{n}(0) =
1)\right]^{2} + q_{1}^{(p)}\,i_{2}(t' \,\vert \boldsymbol{n}(0) =
1),
\]
and
\[
\left(\frac{\displaystyle 1}{\displaystyle
\imath}\right)^{2}\,\left\{\frac{\displaystyle \partial^{2}
q^{(d)}\left[g(\omega, t' \, \vert \boldsymbol{c}(0) =
1)\right]}{\displaystyle \partial \omega^{2}}\right\}_{\omega=0} =
q_{2}^{(d)}\,\left[i_{1}(t' \,\vert \boldsymbol{c}(0) =
1)\right]^{2} + q_{1}^{(d)}\,i_{2}(t' \,\vert \boldsymbol{c}(0) =
1).
\]
Finally, one obtains
\[
\left(\frac{\displaystyle 1}{\displaystyle
\imath}\right)^{2}\,\left[\frac{\displaystyle
\partial^{2} q(\omega, t')}
{\displaystyle \partial \omega^{2}}\right]_{\omega=0} =
q_{2}^{(p)}\,\left[i_{1}(t' \,\vert \boldsymbol{n}(0) =
1)\right]^{2} + q_{1}^{(p)}\,i_{2}(t' \,\vert \boldsymbol{n}(0) = 1)
+ \] \[2 \, q_{1}^{(p)}\, q_{1}^{(d)}\,i_{1}(t' \,\vert
\boldsymbol{n}(0) = 1) \, i_{1}(t' \,\vert \boldsymbol{c}(0) = 1) +
\vspace{0.2cm}
\]
\begin{equation} \label{58}
q_{2}^{(d)}\,\left[i_{1}(t' \,\vert \boldsymbol{c}(0) =
1)\right]^{2} + q_{1}^{(d)}\,i_{2}(t' \,\vert \boldsymbol{c}(0) =
1).
\end{equation}
Before calculating the function $i_{2}(t \,\vert \boldsymbol{n}(0) =
1)$, the parameters \[ q_{1}^{(p)}, \quad q_{2}^{(p)}, \quad
q_{1}^{(d)}, \quad q_{2}^{(d)} \] have to be determined. By using
the generating functions $q^{(p)}(z)$ and $q^{(d)}(z)$ defined by
formulae (\ref{14}), one obtains
\begin{eqnarray}
\left[\frac{\displaystyle d q^{(p)}(z)}{\displaystyle
dz}\right]_{z=1} & = & q_{1}^{(p)} = \nu_{p},  \nonumber \\
\left[\frac{\displaystyle d^{2} q^{(p)}(z)}{\displaystyle
dz^{2}}\right]_{z=1} & = & q_{2}^{(p)} = <\nu_{p}(\nu_{p}-1)>, \nonumber \\
\left[\frac{\displaystyle d q^{(d)}(z)}{\displaystyle
dz}\right]_{z=1} & = & q_{1}^{(d)} = \nu_{d}, \nonumber  \\
\left[\frac{\displaystyle d^{2} q^{(d)}(z)}{\displaystyle
dz^{2}}\right]_{z=1} & = & q_{2}^{(d)} = <\nu_{d}(\nu_{d}-1)>
\nonumber.
\end{eqnarray}
In the next step one substitutes (\ref{58}) into (\ref{55}), which leads to
\[
i_{2}(t \,\vert \boldsymbol{n}(0) = 1) = \lambda_{d}\,\int_{0}^{t}
e^{-\lambda_{r} (t-t')} \, \Phi_{2}(t')\;dt' +
\lambda_{f}\,\int_{0}^{t} e^{-\lambda_{r}
(t-t')}\,\left\{q_{1}^{(p)}\,i_{2}(t' \,\vert
\boldsymbol{n}(0) = 1) + \right.\] \[\left. q_{1}^{(d)}\,i_{2}(t' \,\vert
\boldsymbol{c}(0) = 1) + q_{2}^{(p)}\,\left[i_{1}(t' \,\vert
\boldsymbol{n}(0) = 1)\right]^{2} + \right. \vspace{0.2cm}
\]
\begin{equation} \label{59}
\left. 2 \, q_{1}^{(p)}\, q_{1}^{(d)}\,i_{1}(t'
\,\vert \boldsymbol{n}(0) = 1) \, i_{1}(t' \,\vert \boldsymbol{c}(0)
= 1) + q_{2}^{(d)}\,\left[i_{1}(t' \,\vert
\boldsymbol{c}(0) = 1)\right]^{2} \right\}\;dt'.
\end{equation}
In the further calculations one needs the relations
\begin{equation} \label{60}
i_{j}(t \,\vert \boldsymbol{c}(0) = 1) = \lambda \, \int_{0}^{t}
e^{-\lambda (t-t')}\,i_{j}(t' \,\vert \boldsymbol{n}(0) = 1)\;dt',
\qquad  j=1,2, \ldots,
\end{equation}
which can be obtained from (\ref{24}). For simpler notation it is
useful to introduce the function
\[R(t') = q_{2}^{(p)}\,\left[i_{1}(t' \,\vert
\boldsymbol{n}(0) = 1)\right]^{2} + \]
\begin{equation} \label{61}
2 \, q_{1}^{(p)}\, q_{1}^{(d)}\,i_{1}(t' \,
\vert \boldsymbol{n}(0) = 1) \, i_{1}(t' \,\vert \boldsymbol{c}(0)
= 1) + q_{2}^{(d)}\,\left[i_{1}(t' \,\vert
\boldsymbol{c}(0) = 1)\right]^{2}
\end{equation}
into the integral equation (\ref{59}). Eq. (\ref{59}), can be solved
by Laplace transform. One obtains
\[( s + \lambda_{r})\,\widetilde{i}_{2}(s \,\vert \boldsymbol{n}(0)
= 0) =\]
\begin{equation} \label{62}
\lambda_{d}\,\widetilde{\Phi}_{2}(s) + \lambda_{f} \, \nu_{0}
\,\widetilde{i}_{2}(s \,\vert \boldsymbol{n}(0) = 1) + \lambda_{f}
\, \nu_{d} \,\widetilde{i}_{2}(s \,\vert \boldsymbol{c}(0) = 1) +
\lambda_{f} \, \widetilde{R}(s),
\end{equation}
where
\begin{equation} \label{63}
\widetilde{i}_{2}(s \,\vert \boldsymbol{c}(0) = 1) =
\frac{\displaystyle \lambda}{\displaystyle s + \lambda}\,
\widetilde{i}_{2}(s \,\vert \boldsymbol{n}(0) = 1)
\end{equation}
and
\[\widetilde{R}(s) = \int_{0}^{\infty} e^{-s t}\, R(t)\;dt = \]
\[q_{2}^{(p)}\,\int_{0}^{\infty} e^{-s t}\,\left[i_{1}(t \,\vert
\boldsymbol{n}(0) = 1)\right]^{2} \;dt + 2 \, q_{1}^{(p)}\,
q_{1}^{(d)}\,\int_{0}^{\infty} e^{-s t}\,i_{1}(t \, \vert
\boldsymbol{n}(0) = 1) \, i_{1}(t \,\vert \boldsymbol{c}(0) = 1)\;dt
+ \]
\begin{equation} \label{64}
q_{2}^{(d)}\,\int_{0}^{\infty} e^{-s t}\,\left[i_{1}(t \,\vert
\boldsymbol{c}(0) = 1)\right]^{2}\;dt.
\end{equation}
Accounting for (\ref{63}), after elementary calculations one arrives
at
\begin{equation} \label{65}
\left( s + \lambda_{r} - \lambda_{f} \, \nu_{0} - \lambda_{f} \,
\nu_{d} - \frac{\displaystyle \lambda}{\displaystyle s + \lambda}
\,\right)\,\widetilde{i}_{2}(s \,\vert \boldsymbol{n}(0) = 0) =
\lambda_{d}\,\widetilde{\Phi}_{2}(s) + \lambda_{f} \,
\widetilde{R}(s),
\end{equation}
where
\begin{equation} \label{66}
s + \lambda_{r} - \lambda_{f} \, \nu_{0} - \lambda_{f} \, \nu_{d} -
\frac{\displaystyle \lambda}{\displaystyle s + \lambda} =
\frac{\displaystyle (s + s_{1})\,(s + s_{2})}{\displaystyle s +
\lambda},
\end{equation}
i.e.
\begin{equation} \label{67}
\widetilde{i}_{2}(s \,\vert \boldsymbol{n}(0) = 0) =
\frac{\displaystyle s + \lambda}{\displaystyle (s + s_{1})\,(s +
s_{2})}\,\left[\lambda_{d}\,\widetilde{\Phi}_{2}(s) + \lambda_{f} \,
\widetilde{R}(s)\right]
\end{equation}
By using (\ref{53}) and (\ref{54}), the Laplace transform of the
variance of the detector current $I_{2}(t)$ is obtained as
\begin{equation} \label{68}
\widetilde{I}_{2}(s) = s_{0}\,\frac{\displaystyle (s +
\lambda)}{\displaystyle s\,(s + s_{1})\,(s +
s_{2})}\,\left[\lambda_{d}\,\widetilde{\Phi}_{2}(s) + \lambda_{f} \,
\widetilde{R}(s)\right].
\end{equation}
The calculation of the inverse Laplace transform of
$\widetilde{I}_{2}(s)$ is rather complicated task. It is easier to
calculate it by symbolic manipulation codes. We have used
Mathematica \cite{mathematica} by Wolfram for solving the present
problem.

\subsection{Concrete example for the variance}

The second moment of $\boldsymbol{\eta_{s}}(t)$ with the selected
particular detector pulse shape (\ref{46a}) has the form
\begin{equation} \label{69}
\Phi_{2}(t) = \left[\Phi_{1}(t)\right]^{2} =
\alpha_{e}^{4}\,t^{2}\,e^{-2 \alpha_{e} t}\, Q^{2},
\end{equation}
whose Laplace transform is
\begin{equation} \label{70}
\widetilde{\Phi}_{2}(s) = 2 \, \frac{\displaystyle
\alpha_{e}^{4}}{\displaystyle (s + 2\,\alpha_{e})^{3}}\,Q^{2}.
\end{equation}
Since the expression of
\begin{equation} \label{71}
\widetilde{R}(s) = q_{2}^{(p)}\,\widetilde{R}^{(p)}(s) +
2\,q_{1}^{(p)}\,q_{1}^{(d)}\widetilde{R}^{(0,d)}(s) +
q_{2}^{(d)}\,\widetilde{R}^{(d)}(s)
\end{equation}
is prohibitively long, we do not reproduce it here in print.

\begin{figure}[ht!]
\protect \centering{
\includegraphics[scale=0.8] {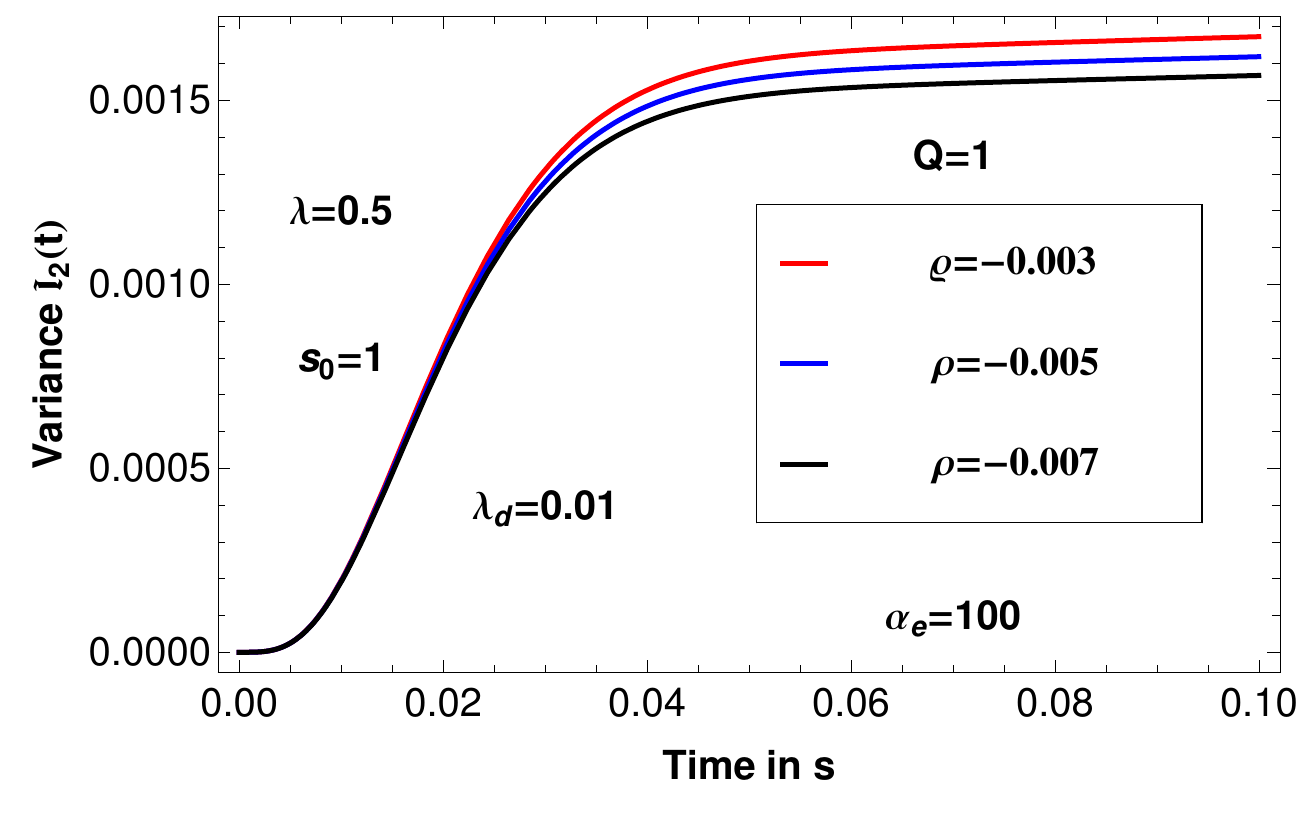}}\protect
\caption{\label{fig5}\footnotesize{Time dependence of the variance
of the detector current $I_{2}(t)$ in a subcritical assembly just
after the neutron source was switched on, at three different
reactivities}}
\end{figure}
In order to illustrate the effect of prompt neutrons just after the
switching on of the neutron source, the time dependence of
$I_{2}(t)$ for small values of $t$ was calculated. Fig. \ref{fig5}
shows the time dependence of the variance $I_{2}(t)$ of the detector
current in a subcritical assembly just after the neutron source was
switched on, for three different reactivities. From this figure it
would appear as if the system, as monitored by the detector signal,
reached the stationary state rather fast. An inspection of the
long-time behaviour of the system shows, however, that this is note
the case, and one has to follow up the behavior of the function
$I_{2}(t)$ during a much longer period to arrive at the stationary
variance of the detector current.

\begin{figure}[ht!]
\protect \centering{
\includegraphics[scale=0.8] {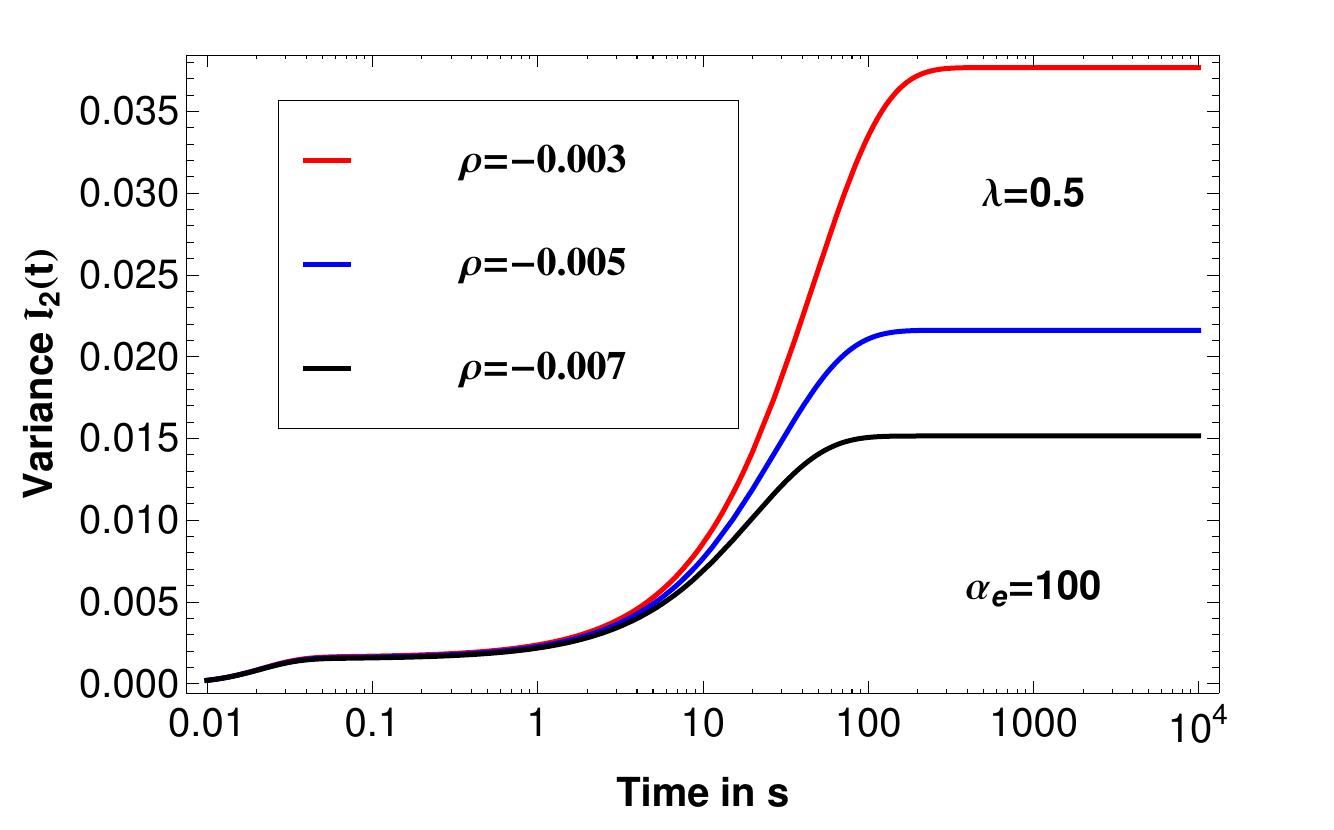}}\protect
\caption{\label{fig6}\footnotesize{Approach to the stationary value
of the variance of the detector current in a subcritical assembly
with a source switched on at $t=0$, for three different
reactivities}}
\end{figure}
Fig. \ref{fig6} shows the approach to the stationary value
$I_{2}^{(st)}$ of the variance of the detector current for three
different reactivities during a longer time period. It is seen how
the presence of delayed neutrons extend significantly the time
needed to reach stationarity.

\begin{figure}[ht!]
\protect \centering{
\includegraphics[scale=0.8] {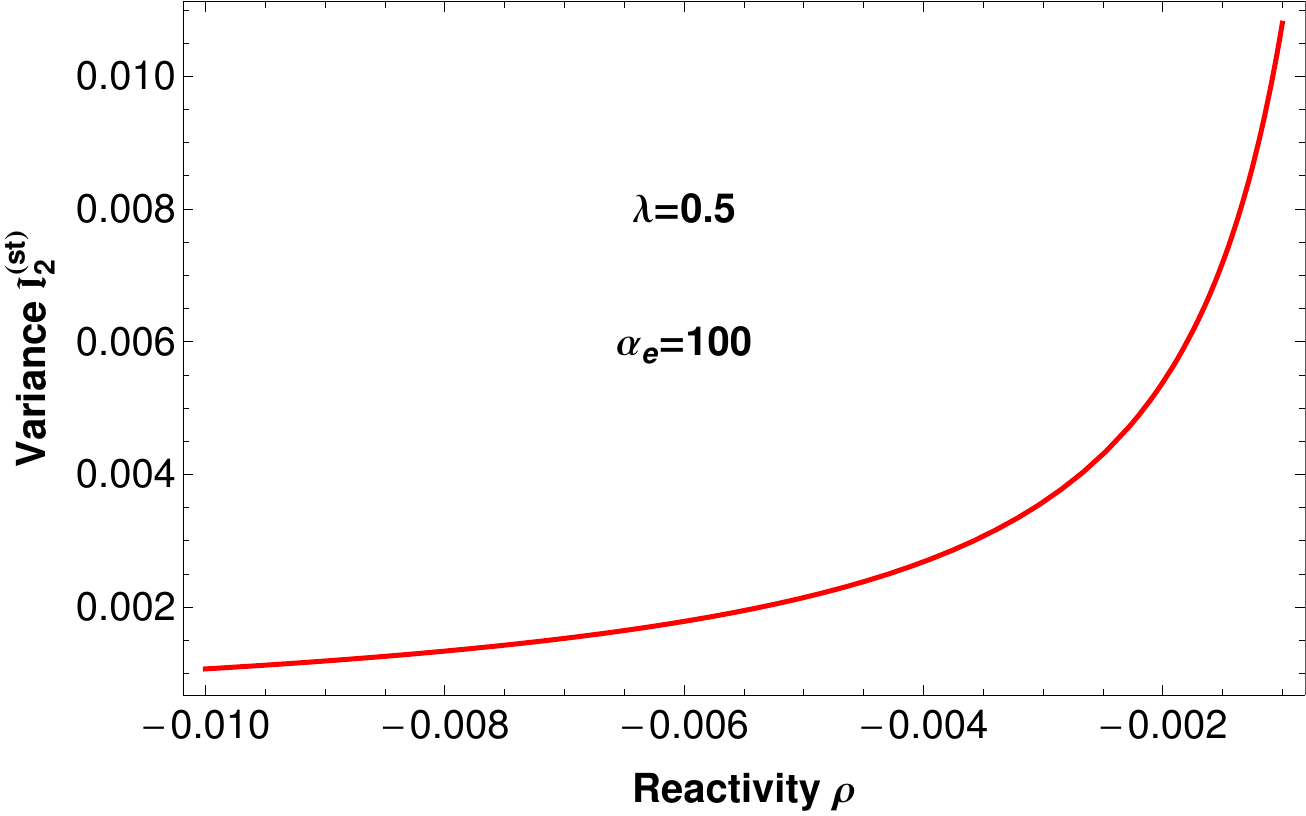}}\protect
\caption{\label{fig7}\footnotesize{Dependence of the stationary
variance of the detector current $I_{2}^{(st)}$ on the reactivity
$\rho$ of a subcritical assembly}}
\end{figure}

In practice one needs the dependence of the stationary variance of
the detector current on the reactivity $\rho$. With the help of the
Tauber theorem, one finds that
\begin{equation} \label{72}
\so{\boldsymbol{\eta}^{(st)}} =
I_{2}^{(st)} = \lim_{s \rightarrow 0} s\, \widetilde{I}_{2}(s) =
s_{0}\,\frac{\displaystyle \lambda_{d}\,\alpha_{e} \,
\lambda}{\displaystyle 4 s_{1}\,s_{2}} \,Q^{2} \,\mathbb{C}(\rho),
\end{equation}
The explicit form of the function $\mathbb{C}(\rho)$ is rather
lengthy, therefore it is given  in the Appendix. The reason for
separating out the multiplying factors in (\ref{72}) from the
function $\mathbb{C}(\rho)$ will be clear in the forthcoming
discussion, where a comparison with the results of the traditional
Campbell theorem will be shown. For an illustration, the dependence
of $I_{2}^{(st)}$ on the reactivity $\rho$ is shown in Fig.
\ref{fig7}.
\begin{figure}[ht!]
\protect \centering{
\includegraphics[scale=0.8] {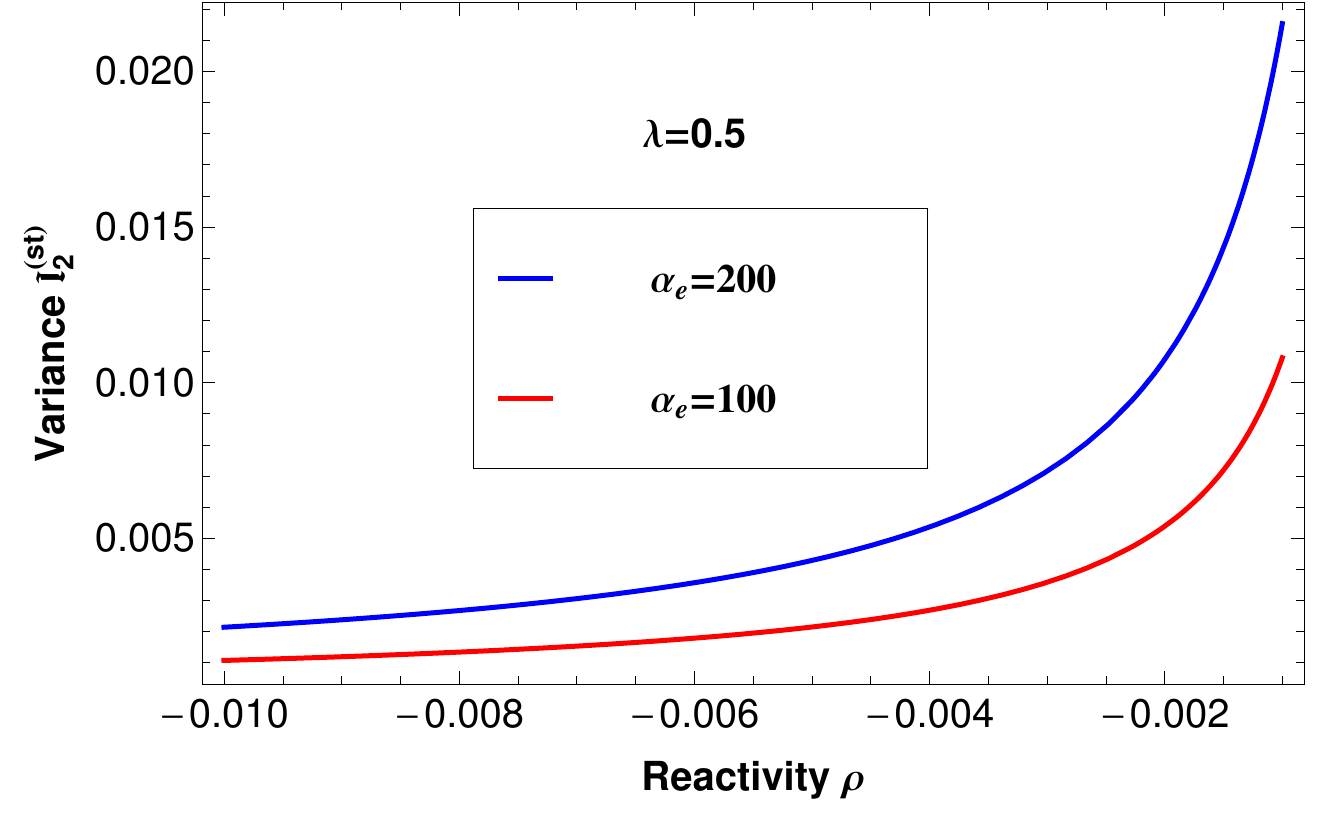}}\protect
\caption{\label{fig8}\footnotesize{Dependence of the stationary
variance of the detector current $I_{2}^{(st)}$ on the reactivity
$\rho$ of a subcritical assembly at two decay constants $\alpha_{e}$
}}
\end{figure}

Fig. \ref{fig8} shows the sensitivity of $I_{2}^{(st)}$ to the
variation of the detector pulse decay constant $\alpha_{e}$. One
finds that a larger $\alpha_{e}$ brings about a larger stationary
variance of the detector current.

\section{Discussion}

In possession of the result for the variance of the detector signal
for the case of detection in a multiplying medium, it is possible to
compare it with the value obtained from the application of the
traditional Campbell formula. Such a comparison was already made in
our previous work \cite{palpaz15}, for the case where only prompt
neutrons were assumed in the fission chain. In addition to
performing the same analysis by accounting for the delayed neutrons,
also some further aspects will be discussed, which were not analysed
in the previous work.

As mentioned in \cite{palpaz15}, for a correct comparison, one has
to account for the fact that the traditional Campbell formulae are
expressed in terms of the intensity of the \textit{detection events}
in the detector (which will be denoted here as $s_{0}^*$), whereas
in the present formulae the intensity $s_{0}$ of the
\textit{injection of neutrons} from an extraneous source appears.
The asterisk here is to indicate that the two intensities are
intrinsically different. In the continuation, in order to
distinguish between the traditional formulae (independent events)
and the present ones (non-independent detection events), the former
will be denoted by an asterisk.

Since in a subcritical medium with reactivity $\rho$ and an
extraneous neutron source with intensity $s_{0}$, the stationary
neutron density is given as\footnote{For a recent note on a general
misconception regarding the derivation of this formula, see Ref.
\cite{paz15let}} $n^{(st)} = s_{0} \Lambda / (-\rho$), the detection
intensity $s_{0}^*$ will be equal to
\begin{equation}\label{73}
s_{0}^* = s_{0} \frac{\lambda_{d} \Lambda }{-\rho}
\end{equation}
For a correct comparison, the traditional Campbell formulae need to
be used with a detection intensity $s_{0}^*$ as above.

The stationary variance of a detector signal with independent
incoming events, $\boldsymbol{\eta}_{*}^{(st)}$, with the detector
response function given by Eq. (\ref{46a}), was already calculated
in Ref. \cite{PPE14} with the result
\begin{equation}\label{74}
\so{\boldsymbol{\eta}_{*}^{(st)}} = \frac{1}{4}s_{0}^{*} \alpha_{e}
Q^{2} = \frac{1}{4} s_{0} \, \frac{\lambda_{d} \, \Lambda
\,\alpha_{e}}{- \rho}Q^{2} = s_{0}\ \frac{\displaystyle
\lambda_{d}\,\alpha_{e} \, \lambda}{\displaystyle 4 s_{1} \,s_{2}}
\,Q^{2}.
\end{equation}
Here, in order to facilitate the comparison with the formula
obtained in this paper for non-independent incoming events, Eq.
(\ref{72}), in the last equality we used the identity $s_{1} s_{2} =
-\lambda\, \rho/\Lambda$, which can easily be obtained from Eqs
(\ref{40}) and (\ref{41}). With $s_{0}^{*}$ properly expressed in
terms of $s_{0}$, the bias of the traditional formula, when using it
for the evaluation of measurements made in a subcritical or a
critical core where the primary detection events are not
independent, can be expressed by the ratio of the correct variance
obtained in the present results, to the variance of the traditional
Campbell method. By using (\ref{72}) and (\ref{74}) one obtains
\begin{equation}\label{75}
\frac{\so{\boldsymbol{\eta}^{(st)}}}{\so{\boldsymbol{\eta}_{*}^{(st)}}}
= \mathbb{C}(\rho)
\end{equation}
where the function $\mathbb{C}(\rho)$, giving the bias of the
traditional formula, is given in the Appendix.

\begin{figure}[ht!]
\protect \centering{
\includegraphics[scale=0.8] {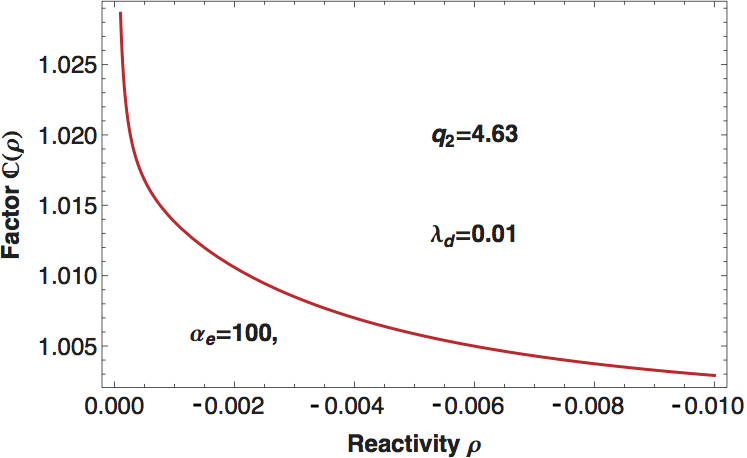}}\protect
\caption{\label{fig10}\footnotesize{Dependence of the factor
$\mathbb{C}(\rho)$ for the stationary variance of the  detector
current on the reactivity}}
\end{figure}

\begin{figure}[ht!]
\protect \centering{
\includegraphics[scale=0.9] {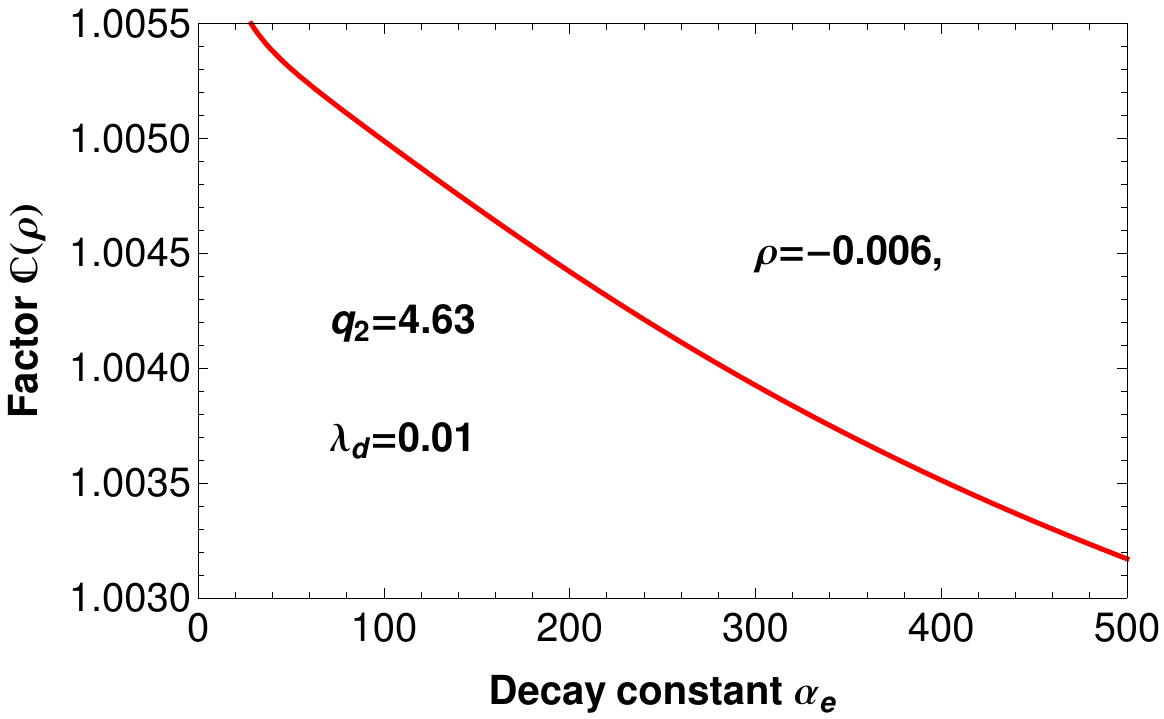}}\protect
\caption{\label{fig11}\footnotesize{Sensitivity of the factor
$\mathbb{C}(\rho)$ to the decay constant $\alpha_{e}$}}
\end{figure}

Figure \ref{fig10}. shows the dependence of the bias factor
$\mathbb{C}(\rho)$ on $\rho$. The sensitivity of the bias factor to
the detector signal decay constant $\alpha_{e}$ can be seen in
Figure \ref{fig11}.

The dependence of $\mathbb{C}(\rho)$ on $\rho$, although
algebraically much more complicated, follows the same tendency as in
the previous case, i.e. without delayed neutrons in the
multiplication process (Ref. \cite{palpaz15}, Fig. 7). It is seen
that, as the system approaches criticality ($(-\rho) \rightarrow
0$), the bias factor diverges. This behaviour can also be readily
derived from Eq. (\ref{76}), by noting from (\ref{40}) that
\begin{equation}
\lim_{(-\rho) \rightarrow 0} s_{1} = \lambda + \frac{\beta}{\Lambda}
\qquad \mbox{and} \qquad s_{2} \rightarrow  - \frac{\displaystyle
\lambda}{\displaystyle \lambda\, \Lambda + \beta}\,\rho
 \quad \mbox{if} \quad -\rho \rightarrow 0.
\end{equation}

Using also the identity
\[
s_{1} s_{2} = -\frac{\lambda\, \rho}{\Lambda}
\]
it is readily seen that for vanishing reactivities,
$\mathbb{C}(\rho)$ diverges as $(-\rho)^{-1}$. This is because the
stationary variance of the neutron number in a critical reactor
diverges, and the correct formula for the variance of the detector
signal mirrors this fact.

From the practical point of view, Eqs (\ref{72}) and (\ref{73}) show
that  in measurements in a subcritical core, the variance of the
detector signal is still proportional to the detection intensity,
which in its turn is proportional to the neutron flux, although with
a proportionality factor which is not given properly by the
traditional Campbell theorem. In addition, the bias of the
traditional Campbell formula depends also on the level of
subcriticality. However, from the quantitative point of view, Fig.
\ref{fig10} shows that even for moderate subcriticalities (i.e.
close to critical, such as $\rho = 0.002 \approx -0.3\,\$$,
corresponding to $k_{eff} = 0.998$), the bias factor is still quite
close to unity. Hence, fission chambers can be used without problems
to measure the stationary flux in subcritical systems, as long as
the detector was properly calibrated in measurements, to obtain the
correct proportionality factor.

The apparent problem that the bias factor diverges for critical
systems, does not constitute a problem in practice either. In
reality, one determines an estimate of the exact value of the
calibration factor, whose definition is based on an ensemble
average, from a time average, which is taken over a finite time
interval, hence it remains finite. This is very much in line with
the way how the auto power spectral densities of time-resolved
measurement signals are determined from the Fourier transforms of
the signal (as opposed to that of its auto-correlation function) by
the help of the Wiener-Khinchin theorem, which also involves the
appearance of a scaling factor which is infinite, but whose estimate
remains a finite quantity.

Interesting insight can be gained by taking the opposite case of
deeply subcritical systems with $\rho \rightarrow -\infty$. Fig.
\ref{fig10} indicates that in that case the bias factor tends to a
constant. A simple analysis of Eq. (\ref{76}) shows that, since
\begin{equation}
s_{1} \rightarrow  - \frac{\rho}{\Lambda} \quad \mbox{if} \quad \rho
\rightarrow - \infty; \qquad \mbox{and} \qquad \lim_{\rho
\rightarrow - \infty} s_{2} = \lambda
\end{equation}
one has \begin{equation}
\lim_{\rho \rightarrow - \infty} \mathbb{C}(\rho) = 1
\end{equation}
This is a completely logical result, which expresses the fact that
in a deeply subcritical system the effect of branching diminishes.
Hence, in the limit, the individual detection events will become
independent, and the traditional formula and the one accounting for
the correlations, give exactly the same result. This shows a nice
and reassuring link between the traditional Campbelling theory and
the more involved case where the non-independent character of the
incoming primary events is accounted for. It is worth mentioning
that the same agreement between the traditional and the extended
theory is found in our previous paper \cite{palpaz15}, i.e. that the
bias factor converges to unity when $\rho \rightarrow - \infty$.
This can be obtained analytically from Eq. (60) of Ref.
\cite{palpaz15}, which readily yields
\begin{equation}
\lim_{\alpha \rightarrow \infty} \mathbb{C}(\alpha) = 1
\end{equation}
However this fact was failed to be mentioned in \cite{palpaz15}.

\section{Conclusions}

The previously introduced formalism for the calculation of the
statistics of the signal of a fission chamber, detecting neutrons in
a multiplying medium and hence experiencing non-independent
detection events, was extended by accounting for delayed neutrons.
The variance of the detector signal was derived and explicitly
calculated with the assumption of a plausible detector response
function. A comparison with the variance of the traditional formula,
given by the Campbell theorem, made it possible to quantify the
dependence of the bias on the subcriticality of the multiplying
system. As expected, the deviation between the two cases vanishes in
the case of deep subcriticalities, since in a non-multiplying medium
only the source neutrons will be detected, which are emitted
independently from each other. When approaching criticality, the
effect of the multiplication and hence that of the non-independent
character of the detections will increasingly dominate, thus the
bias of the traditional Campbelling technique will increase. In
practice, as long as the fission chamber is calibrated from
measurements, this does not pose a problem. Although, according to
the results, the calibration factor depends on the system
subcriticality, our results show that the bias factor is
quantitatively quite close to unity for the regimes in which the
planned subcritical accelerator driven systems are planned to be
operated, hence fission chambers can be used for flux monitoring.
For the case of measurements in a critical system, which is the most
important mode of operation of a fission chamber, the divergence of
the bias, and hence that of the corresponding calibration factor is
handled by estimating the variance of the detector signal from a
measurement of finite time duration.

\section*{Appendix}

The factor $\mathbb{C}(\rho)$ in the stationary variance of the
detector current (\ref{72}) is given by the following formula:
\[ \mathbb{C}(\rho) = \] \[\frac{\displaystyle 1}
{\displaystyle s_{1} \, s_{2}\, \left(s_{1} + s_{2}\right)
\,\left(s_{1} + \alpha_{e}\right)^{2} \,\left(s_{2} +
\alpha_{e}\right)^{2}}\,\left\{ s_{1}^{4} \,s_{2}\,\left(s_{2} +
\alpha_{e}\right)^{2} + s_{1}^{3} \,s_{2}\,\left(s_{2} +
\alpha_{e}\right)^{2}\, \left(s_{2} + 2 \alpha_{e}\right) +
\right.\] \[ \left. 2\,\lambda_{d}\, \lambda_{f}\, \lambda^{2}\,
\left(2\,q_{1}^{(p)}\,q_{1}^{(d)} + q_{2}^{(p)} +
q_{2}^{(d)}\right)\,\alpha_{e}\,\left(s_{2} + \alpha_{e}\right)^{2}
+ \right. \] \[\left. 2 s_{1}^{2}\, s_{2}^{4}\, \alpha_{e} + 5
s_{1}^{2}\, s_{2}^{3}\, \alpha_{e}^{2} + 4 s_{1}^{2}\, s_{2}^{2}\,
\alpha_{e}^{3} + 2\,\lambda_{d}\, \lambda_{f}\, \lambda^{2}\,
\left(2\,q_{1}^{(p)}\,q_{1}^{(d)} + q_{2}^{(p)} +
q_{2}^{(d)}\right)\,\alpha_{e}\,s_{1}^{2} + \right. \]
\begin{equation}\label{76}
\left. s_{1}^{2}\, s_{2}\, \left[\alpha_{e}^{4} + \lambda_{d}\,
\lambda_{f}\, q_{2}^{(p)}\, \alpha_{e}^{2} + \lambda_{d}\,
\lambda_{f}\, \lambda^{2}\, \left(2\,q_{1}^{(p)}\,q_{1}^{(d)} +
q_{2}^{(p)} + q_{2}^{(d)}\, \right)\right] + \right.
\end{equation}
\[ \left. s_{1}\, s_{2}^{4}\, \alpha_{e}^{2} + 2 s_{1}\, s_{2}^{3}\,
\alpha_{e}^{3} + 4\, \lambda_{d}\, \lambda_{f}\, \lambda^{2}\,
s_{1}\,  \left(2\,q_{1}^{(p)}\,q_{1}^{(d)} + q_{2}^{(p)} +
q_{2}^{(d)}\, \right)\, \alpha_{e}^{2} + \right. \]
\[ \left. 2 \lambda_{d}\, \lambda_{f}\, s_{1}\, s_{2}\,
\left[2\,\left(2\, q_{1}^{(p)}\, q_{1}^{(d)} +
q_{2}^{(d)}\right)\,\lambda^{2} + q_{2}^{(p)}\, \left(2 \lambda^{2}
+ \alpha_{e}^{2} \right)\right]\,\alpha_{e}  + \right. \]
\[ \left. s_{1}\, s_{2}^{2}\, \left[\alpha_{e}^{4} + \lambda_{d}\,
\lambda_{f}\, q_{2}^{(p)}\, \alpha_{e}^{2} + \lambda_{d}\,
\lambda_{f}\,\lambda^{2}\, \left(q_{1}^{(p)}\, q_{1}^{(d)} +
q_{2}^{(p)} + q_{2}^{(d)}\right)\right]\right\}, \] where $s_{1}$
and $s_{2}$ are defined by (\ref{40}) and (\ref{41}), respectively.
It is worth noting that the formula depends, among others, on the
detector characteristics (the pulse shape and the corresponding
decay constant $\alpha_{e}$ of the detector), hence it cannot be
considered as universal. However, the qualitative monotonic
behaviour, as well as the asymptotic properties for the cases $\lim
-\rho \rightarrow 0$ and $\lim \rho \rightarrow - \infty$ are
universal, and do not depend on the detector properties. It is only
the speed of the convergence which is dependent on the detector
characteristics.

\section*{References}


\end{document}